\documentclass[aps,twocolumn]{revtex4}
\usepackage{graphicx}

\begin{document}

\title{\LARGE\bf Fermi, Pasta, Ulam  and a
 {\em mysterious} lady}
\bigskip


\author{Thierry DAUXOIS}

\affiliation{Universit\'e de Lyon, CNRS, Laboratoire de Physique,
\'Ecole Normale Sup\'erieure de Lyon, 46 All\'{e}e d'Italie, 69364
Lyon c\'{e}dex 07, France\\
Thierry.Dauxois@ens-lyon.fr}

\begin{abstract}
It is reported that the numerical simulations of the
Fermi-Pasta-Ulam problem were performed by a young lady, Mary
Tsingou. After 50 years of omission, it is time for a proper
recognition of her decisive contribution to the first ever numerical
experiment, central in the solitons and chaos theories, but also one
of the very first out-of-equilibrium statistical mechanics study.
Let us quote from now on the Fermi-Pasta-Ulam-{\em Tsingou} problem.
\end{abstract}

\maketitle

The Fermi-Pasta-Ulam problem~\cite{FPU} was named after the three
scientists who proposed to study how a crystal evolves towards
thermal equilibrium. The idea was to simulate a chain of particles,
linked by a linear interaction but adding also a weak {\em
nonlinear} one. FPU thought that, due to this additional term, the
energy introduced into a single Fourier mode should slowly drift to
the other modes, until the equipartition of energy predicted by
statistical physics is reached. The beginning of the calculation
indeed suggested that this would be the case, but to their great
surprise, after a longer time, almost all the energy was back to the
lowest frequency mode and the initial state seems to be almost
perfectly recovered after this recurrence period. Thus, contrary to
the expectations of the authors, the drift of the energy does not
occur. This highly remarkable result, known as the FPU paradox,
shows that nonlinearity is not enough to guarantee the equipartition
of energy.

Pursuing the solution of the FPU paradox, Zabusky and Kruskal
emphasized ten years later the link between the problem in the
so-called continuum limit and the Korteweg-de Vries
equation~\cite{zabusjykruskal}, known to have spatially localized
solutions. Looking to the problem in real space rather than in
Fourier space, they showed how to solve the paradox in terms of the
dynamics of these localized excitations. It was the birth of the
term {\em solitons}, for these localised (or {\em solit}ary) waves
with properties of particles (explaining the suffix {\em on} as for
electron, boson,...). Consequently, the numerous physical
applications~\cite{dauxoispeyrard} of solitons originates from this
FPU paper.

Another line of thought was developed in parallel. People focused on
the Fourier mode dynamics, looking for non-resonance conditions that
could explain the inefficient energy transfer. No convincing
explanation was found before the discovery of the KAM theorem, which
states that most orbits of slightly perturbed integrable Hamiltonian
systems remain quasi-periodic. If the perturbation is so strong that
nonlinear resonances 'superpose', the FPU recurrence is destroyed
and one obtains a fast convergence to thermal
equilibrium.~\cite{chirikov}

The FPU problem is thus of central importance in the Solitons and
Chaos theories~\cite{Chaos}. This is the reason why, in 2005,
several conferences, articles and seminars have celebrated the 50th
anniversary of the May 1955 publication of the Los Alamos report.
This paper marked indeed a true change in modern science, both
making the birth of a new field, {\em Nonlinear Science}, and
entering in the age of computational science: the problem is indeed
the first landmark in the development of physics computer
simulations.

There was however very few mentions of an intriguing point. On the
first page of the FPU Los Alamos report published in 1955, it is
written,

\noindent \mbox{\em ''Report written by Fermi, Pasta and Ulam.}\\
\mbox{\em Work done by  Fermi, Pasta, Ulam {\tt and Tsingou}''.}

This remark, that Mary Tsingou who took part in the numerical study
is not an author of the report, was always puzzling for scientists
who have read this paper: indeed, it is clear that coding the first
ever numerical experiment on the first computer was not a direct and
immediate task. Consequently, why her contribution has it been
recognized only by two lines of acknowledgements? Moreover, why has
it been impossible until today to pick up her track?

People more deeply involved in the FPU literature have usually also
read the 1972 paper by Tuck and Menzel~\cite{tuckandmenzel}. A careful
reading of the introduction clearly emphasizes that one of the author
of this paper, M. T.  Menzel, was coding the original problem: how can
we solve this paradox?

The obvious solution is that in the name M. T. Menzel, M is for Mary
and T for Tsingou. There is no paradox, this is the same person,
after her wedding!  However, once again, it has been impossible for
decades to pick up her trail. We recently discovered however, that
she is still alive and present in Los Alamos, a couple of miles from
the place where this problem, so important in the past and present
of nonlinear physics~\cite{dauxoispeyrard}, was devised. It is time
for a proper recognition of her work.


Born in October 14th 1928 at Milwaukee, Wisconsin, in a Greek native
family, {\sc Mary Tsingou Menzel} spent her childhood in the US. As
the great depression was taking place in the US, her family moved to
Europe in 1936, where her father had a property in Bulgaria.
However, in June 1940 the American embassy advised them to come back
to US for safety. They pick up the very last American ship that left
Italy. Almost within a week after they landed in New York, Italy
declared war.

She gained a Bachelor of Science in 1951 at the University of
Wisconsin, and a Master in mathematics in 1955 at University of
Michigan. In 1952, following a suggestion by her mathematics
professor, a woman, she applied for a position at Los Alamos
National Laboratory. At that time, women were not encouraged to do
mathematics, but because of the Korean war, there was a shortage of
American young men and staff positions were also proposed to young
women. She was thus hired with a whole group of young people right
out of college, for doing hand calculations.

She was initially assigned to the T1 division (T for Theoretical) at
Los Alamos National Laboratory, led during the war by Rudolph
Peierls and to which the famous spy, Klaus Fuchs, belonged. But she
quickly moved to T7 led by N. Metropolis for working on the first
ever computer, the Maniac I, that no one could program. Together
with Mary Hunt, she was therefore the first programmer to start
exploratory work on it. She remembers it as pretty easy because of
the very limited possibilities of the computer: 1000 words.

They were working primarily on weapons but, in parallel, they
studied other problems like programming chess or studying
fundamental physics' problems. Mary Tsingou mostly interacted with
J. R. Pasta. They were the first ones to do actually graphics on the
computer, when they considered a problem with an explosion and
visualized it on an oscilloscope.

In addition to Pasta, she interacted also with Stan Ulam, but very
little with E. Fermi, at that time professor in Chicago. He was
visiting Los Alamos only for short periods, mostly during the
summer. However, she knew  Nella, Fermi's daughter, much better
because Nella didn't want to stay with her parents during their
visits to Los Alamos. Both early twenties girls were sleeping in the
same dormitory, while Enrico and Laura Fermi were hosted by their
good friends, Stan and Fran\c coise Ulam.

\begin{figure}[h!]
\centerline{
\includegraphics[height=5.0cm]{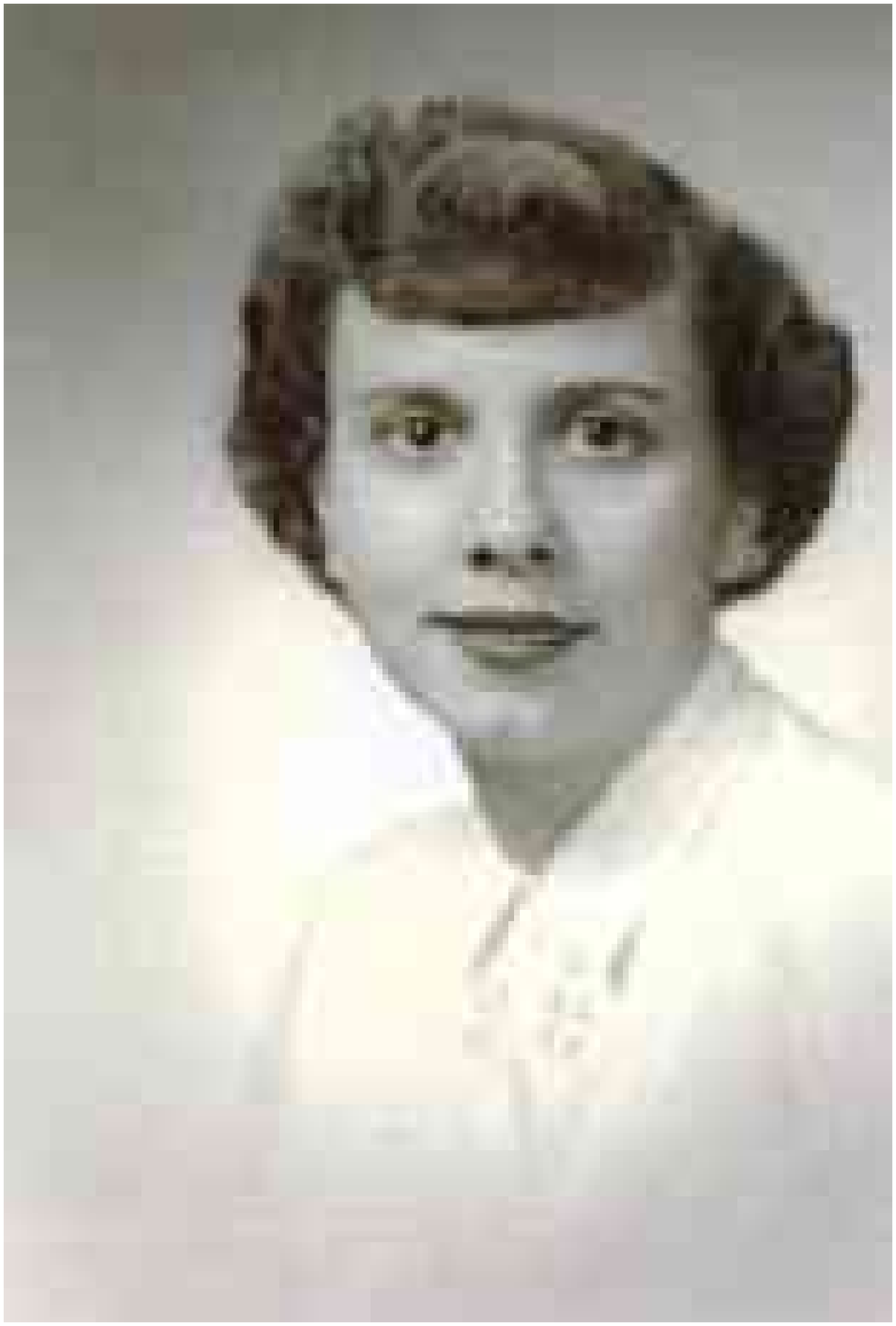}
\includegraphics[height=5.0cm]{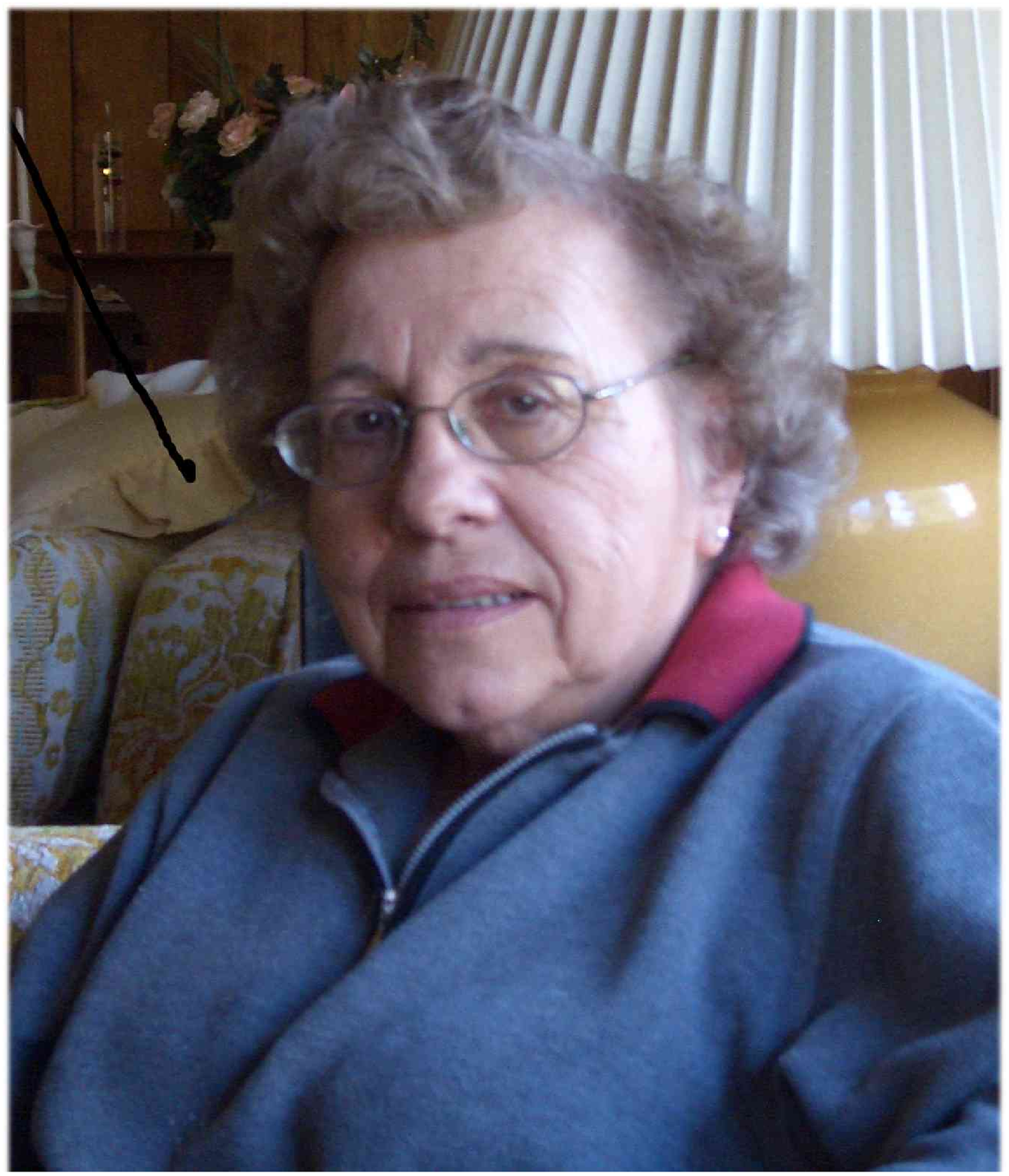}}
\caption{\small Mary Tsingou in 1955 and in 2007. }
\label{fig:marytsingou}
\end{figure}

It was Fermi who had the genius to propose that, instead of simply
performing standard calculus, computers could be used to test a
physical idea, inventing the concept of numerical experiments. He
proposed to check the prediction of statistical physics on the
thermalization of solids. As anticipated, preliminary results
confirmed that energy initially introduced in a single Fourier mode
drift to other ones. However, one day, the oversight to stop the
computer allows one to discover some unexpected recurrences which
were initially hidden by the slowness of the computer. It was the
start of an ongoing fruitful research.~\cite{Chaos}

\begin{figure}[htb]
\centerline{\includegraphics[width=8.5truecm,angle=0]{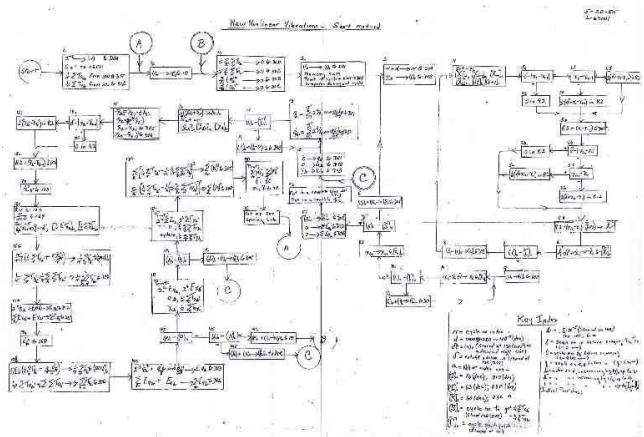}}
\caption{\small Reproduction of the algorithm used by Mary Tsingou
to code the first numerical experiment. Note the date (5-20-55) at
the top right of the figure.} \label{fig:firstprogram}
\end{figure}

The algorithm used by Mary Tsingou in 1955 to simulate the
relaxation of energy in a model crystal on the Maniac is reproduced
in Fig.~\ref{fig:firstprogram}. Its complexity has to be compared
with the 15 lines Matlab$^\copyright$ code, sufficient today to
reproduce the original FPU recurrences~\cite{dauxoispeyrardruffo}.

At the time, programming was a task requiring great insight and
originality, and through the 1960s and even later, it was common to
list programmers as co-authors. It appears that the only reason for
the mention ``Work done by FPU+Tsingou, and report written by FPU"
is that she was not involved in the writing. However, Fermi was not
either since, as noticed in S. Ulam biographical book~\cite{Ulam},
the FPU report was never published because Fermi died before the
writing of the paper. Consequently, Tsingou was not given credit
simply because the report was never formally presented in a journal
and its statement of credit, differentiating between the writing and
the work done, was presumably misread by later people.

In 1958, Mary Tsingou married Joseph Menzel who was also working at
Los Alamos for the Protective Force of the Atomic Energy Commission.
She stayed her whole life in this small city but her colleagues
changed since Metropolis left Los Alamos for Chicago, Pasta went to
Washington, and later Ulam went to Colorado University. She worked
successively on different problems, always with computers. She
became one of the early experts in Fortran (FORmula TRANslator)
invented by IBM in 1955, and was assigned to help researchers in the
laboratory.

After her seminal programming work on the Maniac, in the beginning
of the sixties she came back to the FPU problem with Jim Tuck
looking for recurrences~\cite{tuckandmenzel}.  But she also
considered numerical solution of Schr\"odinger equations, the mixing
problem of two fluids of different densities with J. Von Neumann,
and other problems. Finally, in the eighties during Ronald Reagan's
presidency, she was deeply involved in the Star Wars project
calculations.

Retired in 1991, Mary T. Menzel is still living with her husband at
Los Alamos, very close to the place where the FPU problem was
designed and discovered: it is time for a proper recognition of her
contribution: let us quote from now on the Fermi-Pasta-Ulam-{\em
Tsingou} problem.

\medskip
{\bf Acknowledgements}: I would like to thank T. J. Gammel and J.
Barr{\'e} for helps.

\end{document}